\documentclass[english,aps,prl,amsmath,twocolumn,preprintnumber,superscriptaddress,notitlepage]{revtex4-2}

\usepackage[T1]{fontenc}
\usepackage[utf8]{inputenc}
\makeatletter
\usepackage[normalem]{ulem}
\usepackage{amsfonts}
\usepackage{graphicx}
\graphicspath{ {figures/}{figures/eps/}{figures/pdf/} }
\usepackage{dcolumn}
\usepackage{bm}
\usepackage{amssymb} 

\usepackage{hyperref}
\hypersetup{
 linkcolor=blue,urlcolor=blue,citecolor=blue}

\usepackage{nomencl}                    
\usepackage{float}                      

\usepackage{fancyhdr}                   
\usepackage{url}                        
\usepackage[inactive]{srcltx}		 	
\usepackage{relsize}                    
\usepackage{booktabs}                   
\usepackage{mathrsfs}                   
\usepackage[titletoc]{appendix}			
\usepackage{comment}
\usepackage{dsfont}
\usepackage{amsmath}
\usepackage{mathtools}
\usepackage{mathptmx}
\usepackage{newtxmath}
\usepackage{amssymb}
\usepackage{array}
\usepackage{color}
\usepackage{bm}
\usepackage{subcaption}
\usepackage{wrapfig}
\usepackage[percent]{overpic} 
\usepackage{braket}
\usepackage{feynmp-auto}
\usepackage[compat=1.1.0]{tikz-feynman}
\tikzfeynmanset{warn luatex=false} 
\usepackage{siunitx}
\usepackage{qcircuit}
\DeclarePairedDelimiterXPP\BigO[1]{\mathcal{O}}{\lparen}{\rparen}{}{#1}
\DeclareUnicodeCharacter{2212}{-}
\DeclareUnicodeCharacter{2265}{\ensuremath{\geq}}
\usepackage{tikz}

\renewcommand{\hbar}{\mathchar'26\mkern-9mu h}

\begin{document}


\title{Many-Body Simulations of the Fast Flavor Instability}
\author{Zoha Laraib}
\email{zlaraib@vols.utk.edu}
\affiliation{Department of Physics and Astronomy, The University of Tennessee, Knoxville, Tennessee 37996, USA}
\affiliation{National Center for Computational Sciences, Oak Ridge National Laboratory, Oak Ridge, Tennessee 37830, USA}
\author{Sherwood Richers}
\affiliation{Department of Physics and Astronomy, The University of Tennessee, Knoxville, Tennessee 37996, USA}

\date{\today}

\begin{abstract}

The neutrino fast flavor instability dominates the evolution of neutrino flavor within the engines of core-collapse supernovae and neutron star mergers. However, theoretical models of neutrino flavor change that include many-body quantum correlations can differ starkly from similar mean-field calculations. We demonstrate for the first time that the inhomogeneous fast flavor instability is disrupted by many-body correlations using a novel tensor network framework that allows a continuous transition between mean-field and many-body results by tuning the singular value decomposition cutoff value. Generalizing the forward-scattering Hamiltonian to spatially varying conditions, we demonstrate that the timescale of flavor transformation scales logarithmically with system size, suggesting that many-body effects could occur before mean-field instabilities are able to saturate. Our results have significant implications for astrophysical explosion dynamics, nucleosynthesis, and observable neutrino signatures.

\end{abstract}

\maketitle

Core-collapse supernovae (CCSNe) and neutron star mergers (NSMs) drive much of the element synthesis in the universe, are laboratories for physics in extreme environments, and promise insights from gravitational wave and neutrino observations of future nearby events (e.g., \cite{janka_ExplosionMechanismsCoreCollapse_2012,mezzacappa_CoreCollapseSupernova_2023,foucart2023neutrino,burrows2021core}). The vast number of neutrinos in these dense environments drive explosions and alter the neutron-to-proton ratio that determines the outcome of nucleosynthesis in the ejected matter~\cite{Qian1993,wang2022neutrinos,fischer2024neutrinos}.

While vacuum oscillations and coherent forward scattering of neutrinos on background matter drive flavor evolution in the Sun and terrestrial experiments \cite{bahcall_StandardSolarModels_1992,wolfenstein2018neutrino,mikheyev1989resonant,bahcall_StandardSolarModels_1992,suzuki_HistoryNeutrinoTelescope_2009,orebigann_FutureSolarNeutrinos_2021}, 
neutrino-neutrino interactions in the extreme neutrino densities in a supernova produce a number of nonlinear collective flavor phenomena that are not yet well understood, including flavor synchronization, spectral splits, and violent flavor instabilities ~\cite{pantaleone1992neutrino,duan2010collective,tamborra2021new,capozzi2022neutrino,richers_FastFlavorTransformations_2022,volpe2024neutrinos,johns2025neutrino}. 
Although the Mikheyev-Smirnov-Wolfenstein mechanism and so-called "collective" oscillations emerge too far out to directly influence fluid dynamics \cite{duan2011self,dasgupta2012role}, the fast flavor instability (FFI) is thought to occur deep under the shock in regions inaccessible to other mechanisms \cite{nagakura2021and,froustey2024linear,akaho2024collisional,nagakura2025neutrino}. The FFI  arises from differing angular distributions of neutrinos and antineutrinos such that their angular distributions are equal in some directions, independent of neutrino masses or mixing angles~\cite{sawyer2005speed, sawyer2009multiangle,banerjee2011linearized,sarikas2012spurious,chakraborty2016self,dasgupta2017fast}, with significant implications for nucleosynthesis and explosion outcomes~\cite{wu2017imprints,li2021neutrino,just2022fast,fernandez_FastFlavorInstability_2022a,ehring2023fast,wang2025effect}. These processes can substantially reshape the flavor content of the neutrino field and thus impact astrophysical explosion mechanisms, chemical evolution of the universe, and interpretation of the physics at extreme densities that give rise to future observations.

Despite this progress, it is still currently unknown whether these mean-field (MF, i.e., ignoring multi-particle entanglement) phenomena follow from a more fundamental many-body (MB) theory of neutrino flavor evolution. In light of this, several works have modeled distributions of neutrinos as a collection of interacting spin-like degrees of freedom. These approaches go beyond the MF limit and capture quantum correlations, decoherence, and entanglement dynamics~\cite{bell2003speed,friedland2003many,balantekin2006neutrino,patwardhan_ManybodyCollectiveNeutrino_2023} that may affect both nucleosynthesis outcomes and detectable neutrino signals \cite{Cervia2020,patwardhan2021spectral,balantekin2024collective,siwach2025exploring}. While MF calculations have shown converged flavor transformation behavior in the "thermodynamic" limit of many particles, their MB analogues are restricted to relatively small system sizes (typically \( N \lesssim 24000 \), depending on approximations and the amount of entanglement) due to the exponential scaling of computational complexity
\cite{roggero2021dynamical,roggero2021entanglement,roggero2022entanglement,martin2023equilibration,illa2023multi,bhaskar2024timescales}. These MB simulations often rely on idealized assumptions such as isotropy, homogeneity, and forward scattering to be computationally tractable.
Depending on the assumptions and initial conditions, dynamics can scale either logarithmically with the number of neutrinos, implying that they would persist at large neutrino numbers, or polynomially, implying they would be relevant only on very long timescales \cite{roggero2021dynamical,martin2022classical}.
Some studies have also expressed caution regarding the use of interacting plane waves or neutrino beams in many-body neutrino oscillation treatments and have proposed modifications to this approach, such as finite neutrino interaction length/times\cite{shalgar2023we,kost2024once},  momentum-changing (non-forward) neutrino scattering \cite{johns2023neutrino,cirigliano2024neutrino}, and Pauli blocking~\cite{goimil2025pauli}.  An alternative framework for describing quantum correlations in collective oscillations is based on the Bogoliubov-BornGreen-Kirkwood-Yvon (BBGKY) hierarchy \cite{volpe2013extended}. In simplified setups like homogeneous two-beam systems, tensor network methods have enabled tractable MB simulations of a larger number of sites\cite{roggero2021dynamical,Cervia_2022,Siwach_2023}, but extending these techniques to inhomogeneous, anisotropic, and dynamically evolving configurations remains a key open challenge.

A major shortcoming of existing many-body simulations is the assumption of spatial homogeneity. Although MF studies have shown that even small inhomogeneities can qualitatively alter the flavor dynamics by triggering FFIs or breaking coherent structures~\cite{Shalgar2021breaking,fiorillo2024fast, fiorillo2024inhomogeneous}, the large number of required degrees of freedom required for converged inhomogeneous calculations is computationally challenging. However, initial tests with a small number of sites suggest important MB effects if the forward-scattering Hamiltonian approximates the full Hamiltonian well \cite{Shalgar2021invalidating}.

In this work, we present a new many-body framework that explicitly incorporates spatial inhomogeneity into the quantum flavor evolution of a dense neutrino ensemble. By extending the standard forward-scattering spin Hamiltonian to include position-dependent initial flavor distributions and utilizing tensor network methods for efficient simulation, we demonstrate that many-body effects manifest in large system sizes before the onset of the FFI. Our results represent the first systematic treatment of many-body inhomogeneous flavor evolution in astrophysical environments.

\textit{Methods --}
We start by looking at the solution of the time-dependent many-body Schrodinger equation for a general quantum state
\begin{equation}
    |\psi\rangle = \sum_{\sigma_1,...,\sigma_{N_\mathrm{sites}}} g_{\sigma_1,...,\sigma_{N_\mathrm{sites}}} |\sigma_1,...,\sigma_{N_\mathrm{sites}}\rangle
    \label{eq:2.2 TN}
\end{equation}
consisting of $N_\mathrm{sites}$ local, spin-like degrees of freedom $\sigma_i \in \{\uparrow,\downarrow\}$, where $|\uparrow \rangle $ represents an electron flavor state and $|\downarrow \rangle$ represents a muon flavor state. We assume two flavors throughout this work. This state is fully defined by the rank $N_\mathrm{sites}$ tensor with complex components $g_{\sigma_1,...,\sigma_{N_\mathrm{site}}}$. In order to efficiently evolve states with small entanglement entropy, we use the ITensor and ITensorMPS \cite{fishman2022itensor,fishman2022codebase} libraries in Julia. These libraries decompose the tensor $g_{\sigma_1,...,\sigma_{N_\mathrm{site}}}$ into a matrix product state (MPS) \cite{vidal2003efficient,vidal2004efficient,verstraete2008matrix,ran_TensorNetworkContractions_2020} and use truncated singular-value decompositions (SVDs) to compress the quantum state \cite{schollwock2011density,paeckel2019time,roggero2021entanglement}. The cutoff ($c$) in a SVD determines the threshold below which singular values are discarded when connecting two sites in a tensor network. The chosen cutoff selects singular values based on their magnitude, preserving only those above the specified tolerance. Specifically, setting $c=1$ retains only a single singular value during truncation, yielding a product state that reproduces the mean-field approximation, while $c=0$ keeps all singular values, allowing full many-body correlations to develop between sites. While this adaptive truncation can significantly reduce computational costs, it may also discard important information about the quantum state if the cutoff is too large~\cite{roggero2021entanglement,vidal2003efficient}.
consisting of $N_\mathrm{sites}$ local, spin-like degrees of freedom $\sigma_i \in \{\uparrow,\downarrow\}$, where $|\uparrow \rangle $ represents an electron flavor state and $|\downarrow \rangle$ represents a muon flavor state. We assume two flavors throughout this work. This state is fully defined by the rank $N_\mathrm{sites}$ tensor with complex components $g_{\sigma_1,...,\sigma_{N_\mathrm{site}}}$. In order to efficiently evolve states with small entanglement entropy, we use the ITensor and ITensorMPS \cite{fishman2022itensor,fishman2022codebase} libraries in Julia. These libraries decompose the tensor $g_{\sigma_1,...,\sigma_{N_\mathrm{site}}}$ into a matrix product state (MPS) \cite{vidal2003efficient,vidal2004efficient,verstraete2008matrix,ran_TensorNetworkContractions_2020} and use truncated singular-value decompositions (SVDs) to compress the quantum state \cite{schollwock2011density,paeckel2019time,roggero2021entanglement}. The cutoff ($c$) in a SVD determines the threshold below which singular values are discarded when connecting two sites in a tensor network. The chosen cutoff selects singular values based on their magnitude, preserving only those above the specified tolerance. Specifically, setting $c=1$ retains only a single singular value during truncation, yielding a product state that reproduces the mean-field approximation, while $c=0$ keeps all singular values, allowing full many-body correlations to develop between sites. While this adaptive truncation can significantly reduce computational costs, it may also discard important information about the quantum state if the cutoff is too large~\cite{roggero2021entanglement,vidal2003efficient}.

We evolve the many-body quantum state using a time-dependent Hamiltonian
\begin{equation}
    \frac{\partial \psi}{\partial t} = -i H[X(t)] \psi \label{eq:Schrodinger eq}
\end{equation}
where $X(t) = \{\vec{x}_i\}(t)$ is the set of time-varying positions of each site indexed by $i$. We assume that each site represents a cubic volume of side length $w$ containing $N$ neutrinos with number density $ n = N / (w^3)$, each of which has momentum $\vec{p}_i$. The position of each site evolves as $d\vec{x}_i/dt = \mathfrak{c}\hat{p}_i$, where $\mathfrak{c}$ is the speed of light and $\hat{p}_i=\vec{p}_i/|\vec{p}_i|$.

The neutrino self-interaction Hamiltonian is given by  
\begin{equation}
H[X(t)] = \sqrt{2}G_F n\sum_{i<j} \vec{\sigma_i} \cdot \vec{\sigma_j} J_{ij} S(\xi_{ij})\,\,,\label{eq:self-int}
\end{equation}
where the subscripts i and j are site indices. The directional dependence of the neutrino-neutrino interaction is encoded in the geometric factor
\(J_{ij}=(1 - \hat{p}_i \cdot \hat{p}_j) \). In addition, we account for spatial inhomogeneity by introducing a shape function \( S(\xi_{ij}) =(1-|\xi|)\Theta(1-|\xi|)\), where the input \(\xi_{ij} = (x_i - x_j)/ w \) is the distance between two sites normalized by the size of the spatial extent of the site, and $\Theta$ is the Heaviside step function. This function enforces the local nature of the neutrino-neutrino coupling, only allowing sites to interact if they are near each other. While the inter-site distances could in principle be generalized to fully three-dimensional positions $\vec{x_i}$, we restrict the neutrino motion to a single spatial dimension in our simulations. This simplification reduces computational cost while retaining key features of inhomogeneous flavor evolution as a first-case study. This treatment of inhomogeneity is not dissimilar from that of \cite{shalgar2023we}, though our parameterization allows for a straightforward separation of computational and physical parameters and extrapolation to large $N_\mathrm{sites}$.

We use periodic boundary conditions to approximately simulate environments that are homogeneous on scales larger than the domain size $L$ and to connect with prior work in mean-field flavor instability. To do so, we identify the position $x$ with the positions $x\pm L$, such that when a particle leaves one side of the domain it appears on the other side. Local interactions also wrap around the domain.
   
\begin{figure}
    \centering
    \includegraphics[width=\linewidth]{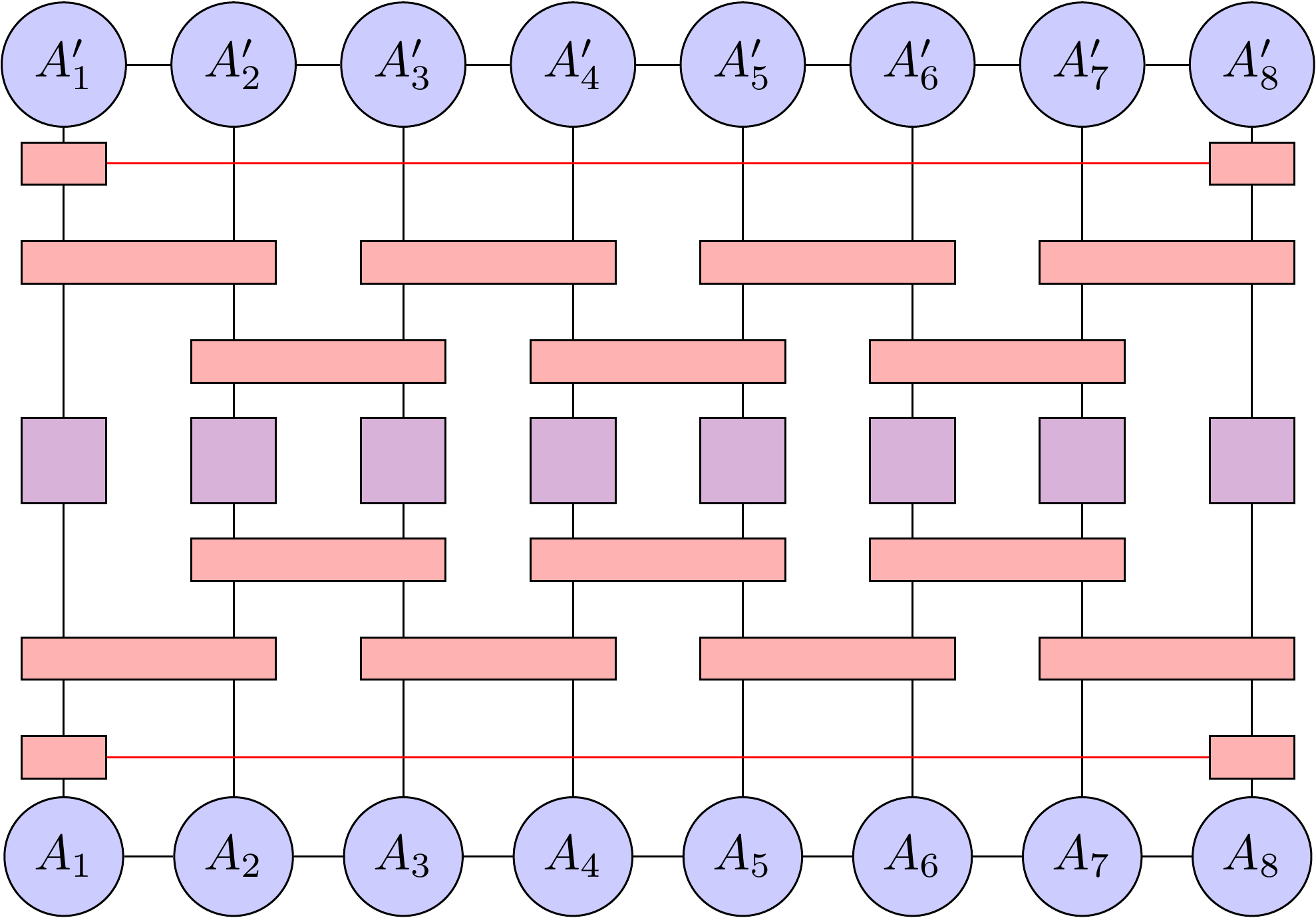}

      \caption{Illustration of operator ordering for an evolution timestep, shown for an example system with 8 sites. Nearest-neighbor gates (salmon) couple adjacent sites, requiring an SVD-based truncation step after their application to control the bond dimension. Purple boxes represent single-site gates that act locally and do not modify the cutoff at each tensor site.}
    \label{fig:operator_order}
\end{figure}
We use time-evolving block decimation (TEBD) to approximate the time evolution operator \cite{vidal2004efficient} using a second-order Trotter-Suzuki decomposition. That is, for a time step of size $\delta$, we apply operators in the forward then reverse order:
\begin{equation}
     \psi(t+\delta) = \prod_{\alpha=1}^{\mathcal{N}_\mathrm{gates}} e^{-i \hat{H}_\alpha \delta /2} \prod^{1}_{\alpha=\mathcal{N}_\mathrm{gates}} e^{-i \hat{H}_\alpha \delta /2}\psi(t)\,\,.
\end{equation}
where $\hat{H}_\alpha$ is an individual one- or two-site gate. We sort sites on the MPS according to their physical positions $x_i$ before each timestep such that a given site will interact with sites to the left and the right (except for the operator that acts on the first and last site to treat the periodic boundary conditions). This operator ordering is shown in Fig.~\ref{fig:operator_order} for an example with 8 sites. Single-site gates act on the site and do not incur any truncation error. Nearest neighbor operators are implemented on a pair of sites, and necessitate reconstructing the truncated SVD after application of the operator.

\textit{Initial Conditions --} Although the FFI occurs in neutrino distributions with nontrivial directional structure, we will simulate a two-beam geometry, as this is the simplest setup that produces the features characteristic of a FFI (i.e., analytically predictable wavelength and growth rate of growing modes, saturation, and equilibration). In order to engineer a mean-field FFI with a fastest growing mode with wavelength $\lambda=1\,\mathrm{cm}$, we set the neutrino number densities as:
\( n = 4.89 \times 10^{32} \, \text{cm}^{-3}
\) (see Equation 2.7 in \cite{chakraborty2016self2}), with a corresponding self-interaction timescale of $\mu^{-1}=(\sqrt{2} G_F n)^{-1} = 10.6\,\mathrm{ps}$. We place an evenly-spaced ($\Delta x = L / (N_{\text{sites}}/2)$) train of electron neutrino moving in the $\hat{x}$ direction, and another superposed evenly spaced train of muon neutrinos moving in the $-\hat{x}$ direction. The particle interactions are determined by a shape function of width $w=\Delta x$, and follow advection in a periodic box of size $L=1\,\mathrm{cm}$. To isolate the fastest-growing mode, we perturb the initial conditions using \( P_x = \pm10^{-6} \sin(kx)\), where initially electron (muon) neutrinos are positively (negatively) perturbed. We use the polarization vectors $P_{j}^i= \langle\psi |\sigma^i_{j}| \psi \rangle$ to represent the full flavor state on all i sites, where $j \in \{x,y,z\}$ indicates the component of the spin operator. This configuration has a mean-field growth rate of $\mathrm{Im}(\omega)=2\mu=1.88 \times 10^{11}\,\mathrm{s}^{-1}$, or a growth timescale of $2\pi/\mathrm{Im}(\omega)=33\,\mathrm{ps}$.

\begin{figure}
    \centering
    \includegraphics[width=\linewidth]{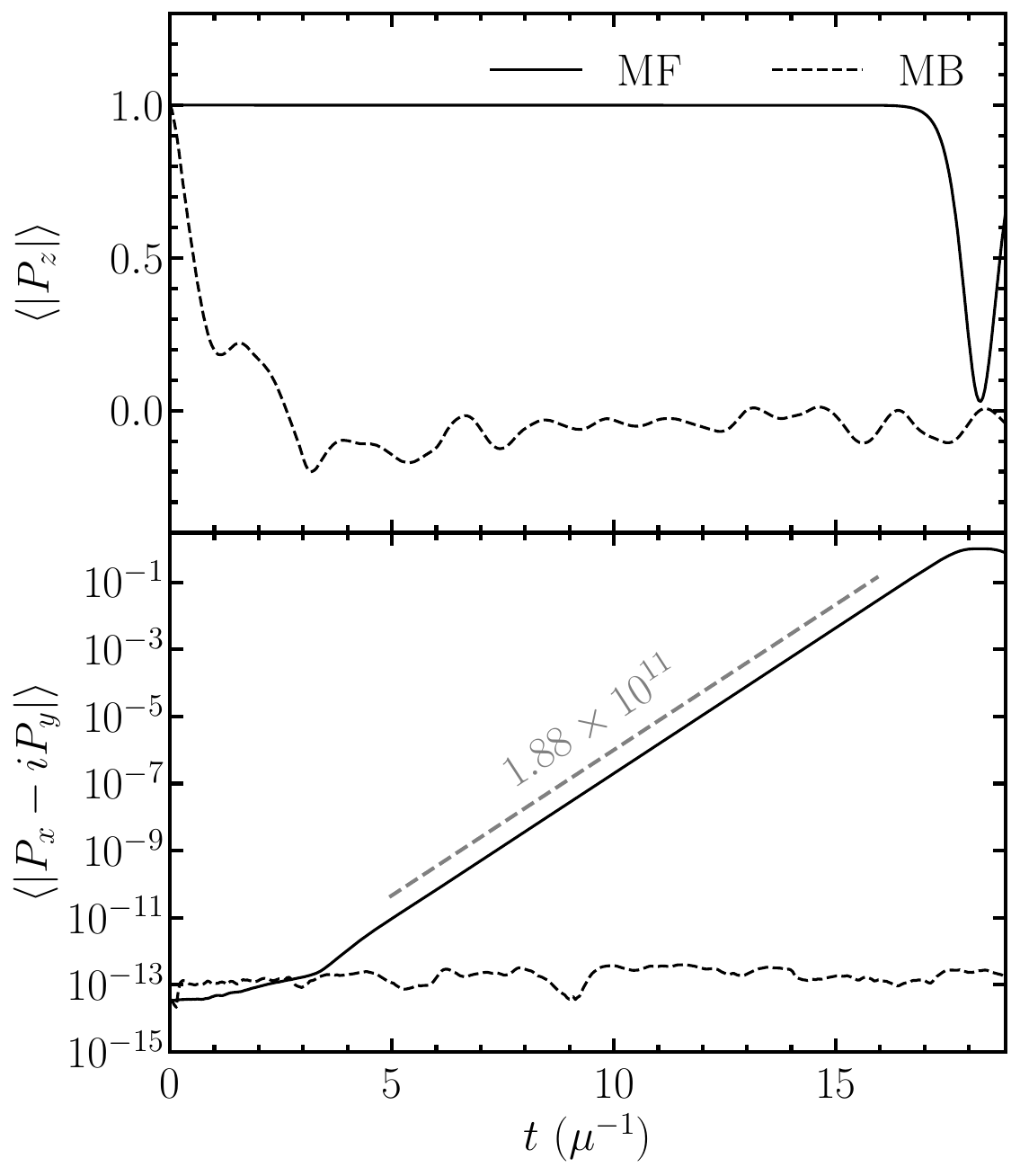}

    \caption{Tensor network simulations of the inhomogeneous two-flavor two-beam fast flavor instability. Opposing neutrino beams each have 10 computational particles initially in each electron (right-moving) and muon (left-moving) flavor states. Mean-field results (solid curve) result from truncating the SVD using a cutoff of $c=1$ and exhibit a FFI growth rate that matches the analytic solution within $0.2\%$. The dashed curve is from an exact (i.e., $c=0$) calculation. Many-body entanglement effects occur before development of the FFI.}
    \label{fig:Richers_Inhomo}
\end{figure}
\textit{Results --} Setting $c=1$ prevents inter-particle entanglement and causes the Hamiltonian to reduce to the well-known mean-field Hamiltonian. We carry out such a mean-field simulation of the FFI with $N_\mathrm{sites}=20$ and demonstrate that we reproduce the instability growth rate predicted by linear stability analysis \cite{chakraborty2016self2} (solid curve in the bottom panel of Fig.~\ref{fig:Richers_Inhomo}) and the amplitude of flavor transformation demonstrated in prior mean-field simulations (solid curve in the top panel of Fig.~\ref{fig:Richers_Inhomo}). The results of the corresponding many-body calculation are shown in Fig.~\ref{fig:Richers_Inhomo}. Instead of amplifying the perturbation exponentially, the neutrinos rapidly undergo flavor transformation from their initial state inducing the system towards depolarization, as seen in the top panel of Fig.~\ref{fig:Richers_Inhomo}. This rapid dip in $P_z$ occurs even before the onset of mean-field (MF) FFI, as shown in the bottom panel. However, the bottom panel also indicates that while flavor transformation develops in the many-body (MB) system, single-site flavor coherence does not.

\begin{figure}
    \centering
        \includegraphics[width=\linewidth]{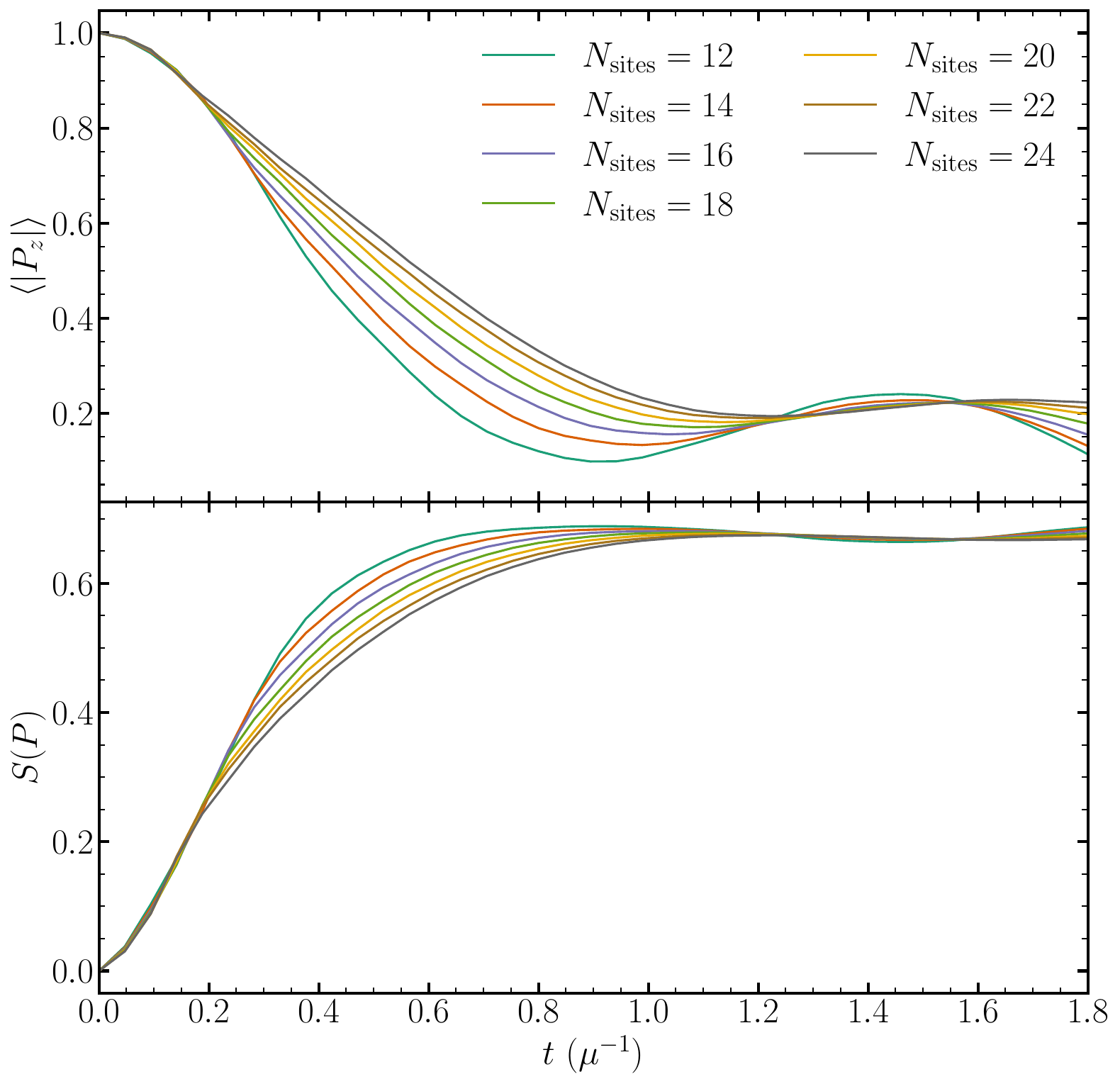}
    \caption{Time evolution of the average magnitude of the $z$-component of the polarization vector $\langle|P_z|\rangle$ (top panel) and the Von-Neumann entanglement entropy $S$ averaged over all sites (bottom panel), computed without truncating the bond dimension for varying system sizes $N_\mathrm{sites}$. Increasing $N_\mathrm{sites}$ delays the onset of significant flavor transformation, shifting both the time of the first minimum ($t_\mathrm{min}$) and the peak entropy to later times.}
    \label{fig:BDinf}
\end{figure}
One may hope that this rapid flavor transformation is an artifact of the small number of computational sites that are used to represent a distribution of many physical neutrinos. We scale $N_\mathrm{sites}$ without adjusting any other parameters to extrapolate the dynamics to the limit of large particle number. In Fig.~\ref{fig:BDinf} we show the evolution of $P_z$ and the Von-Neumann entanglement entropy 
\begin{equation}
S_i = -\frac{1 - P_i}{2} \log\left( \frac{1 - P_i}{2} \right) - \frac{1 + P_i}{2} \log\left( \frac{1 + P_i}{2} \right)\,\,, \label{eq:entropy}
\end{equation}
averaged over all sites. Generally, as $N_\mathrm{sites}$ increases, flavor transformation and entanglement entropy buildup are slower, the value of $\langle|P_z|\rangle$ at the first minimum increases (less overall transformation), and the corresponding peak value of the entanglement entropy decreases. 

We focus on the time of the first minimum in the $\langle|P_z|\rangle$ curve, not for any particular physical significance, but because it is a quantity we can attempt to extrapolate to large $N_\mathrm{sites}$ and compare with the timescale of the FFI onset. In addition, with our chosen domain size and periodic boundary conditions, a particle loops around the domain in $3.14\,\mu^{-1}$, and therefore each pair of particles interacts again on the opposite side of the domain after $1.57\mu^{-1}$. We verify that the properties of the first dip are not affected by domain size using simulations with domain sizes of $L=1\,\mathrm{cm}$, $L=2\,\mathrm{cm}$, and $L=3\,\mathrm{cm}$.

\begin{figure}
    \centering
        \includegraphics[width=\linewidth]{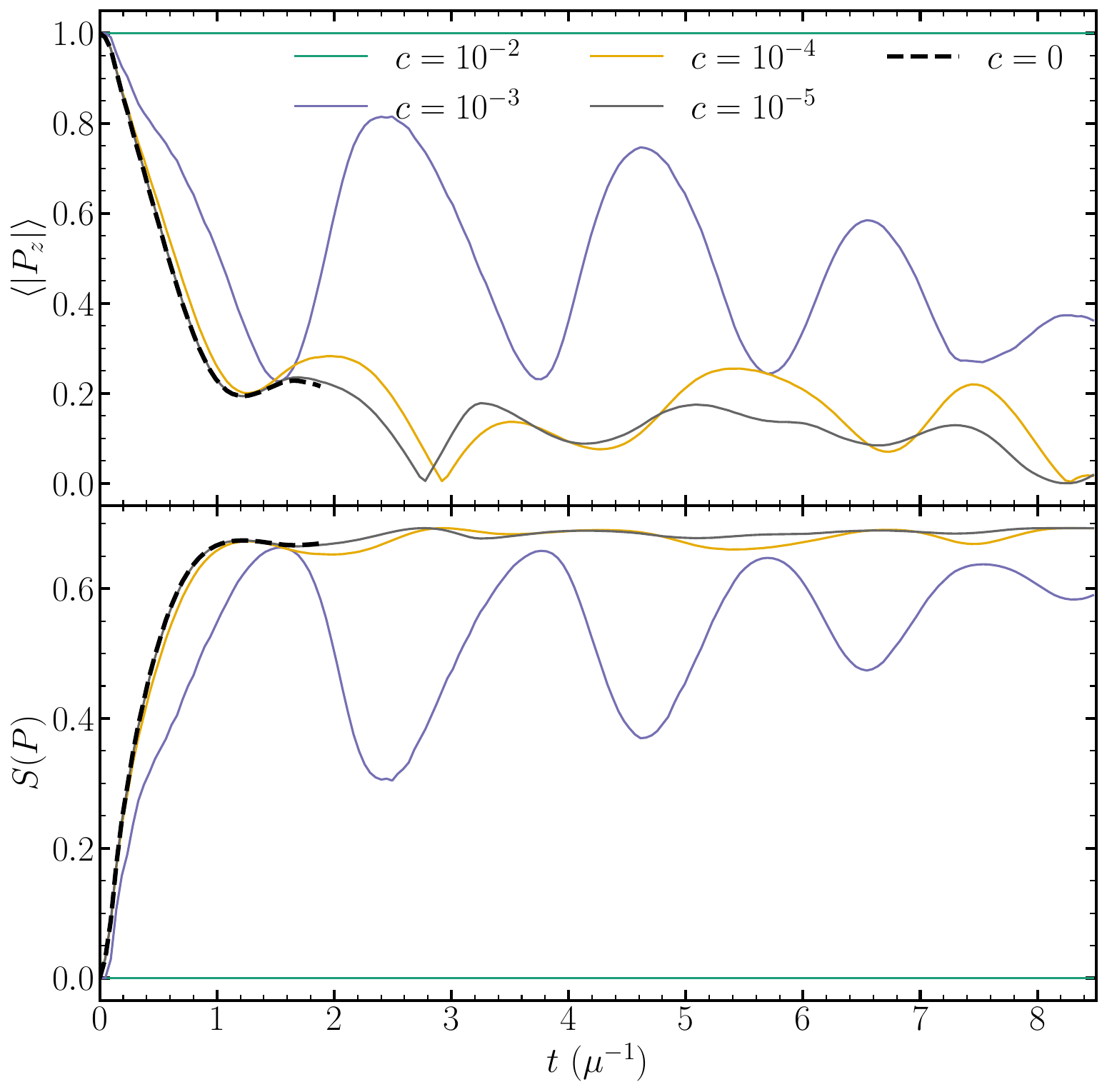}
\caption{Time evolution of the polarization component $\langle|P_z|\rangle$ (top panel) and Von-Neumann entanglement entropy (bottom panel) for $N_\mathrm{sites}=24$, comparing fully-entangled (exact) simulations (black dashed line) with truncated simulations at varying singular-value decomposition thresholds $c$. Larger cutoff values (blue) result in mean-field behavior. Due to the large errors in the $c=10^{-2}$ and $10^{-3}$ calculations, we restrict our later analysis to $c=10^{-4}$, $10^{-5}$, and 0 as given in Table~\ref{tab:fit_parameters} and Fig.~\ref{fig:extrapolate}} .

    \label{fig:tolerance}
\end{figure}
We were able to run fully-entangled simulations with up to $N_\mathrm{sites}=24$. In order to assess scaling to large $N_\mathrm{sites}$, we perform simulations with the same parameters, but approximating the quantum state by truncating SVDs to maintain diagonal entries larger than a chosen cutoff $c$. Fig.~\ref{fig:tolerance} shows the evolution of $\langle|P_z|\rangle$ and entanglement entropy using $N_\mathrm{sites}=24$. Small cutoffs ($c \lesssim 10^{-5}$, red) preserve important singular values, reproducing the full many-body behavior consistent with exact ($c=0$, dashed black) simulations. $c = 10^{-4}$ (orange) remains reasonable, but larger cutoffs ($c = 10^{-3}$, $10^{-2}$) discard smaller yet non-negligible singular values, suppressing weak but important correlations. This artificial loss of entanglement accelerates the transition to mean-field behavior, producing significant deviations from exact many-body dynamics. The excellent agreement at $c=10^{-5}$ justifies its use as a balance between accuracy and computational cost up to at least $N_\mathrm{sites}=24$, though we do not have evidence that this is the relevant cutoff for any other system.

\begin{figure}
    \centering
        \includegraphics[width=1.0\linewidth]{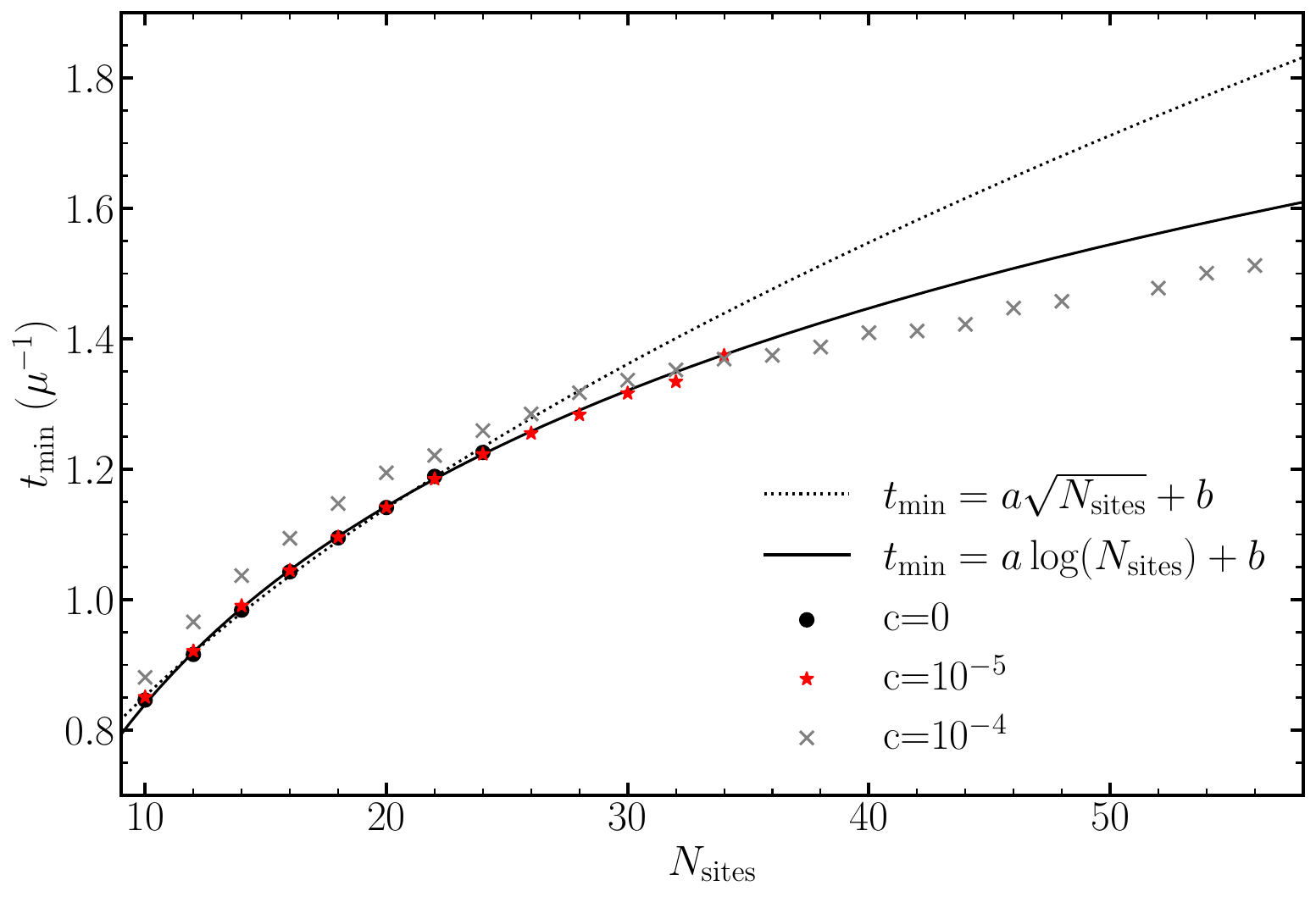}
    \caption{Scaling behavior of the first minimum time $t_\mathrm{min}$ of the polarization component $\langle |P_z|\rangle$ with system size $N_\mathrm{sites}$, computed for different cutoff values: $c=0$ (black dots), $c=10^{-5}$ (red stars), and $c=10^{-4}$ (gray crosses). The curves represent fits to the $c=0$ data using a logarithmic (solid curve) and square root (dotted curve) function. The logarithmic function consistently provides a superior fit for the exact data, with RMS errors significantly lower (see Table~\ref{tab:fit_parameters}).}
    \label{fig:extrapolate}
\end{figure}
\begin{table}[]
    \centering
    \begin{tabular}{c|cc|ccc}
       Function  & $a$ ($\mu^{-1}$) & $b$  ($\mu^{-1}$) & $\epsilon_{c=0}$ & $\epsilon_{c=10^{-5}}$ & $\epsilon_{c=10^{-4}}$ \\\hline
       $a\sqrt{N_\mathrm{sites}}+b$  & $0.220 \pm 0.004$ & \phantom{$-$}$0.157 \pm 0.015$ & 0.005 & 0.032 & 0.141\\
       $a\mathrm{log}(N_\mathrm{sites})+b$ & $0.438\pm0.005$ & $-0.169\pm0.014$ & 0.004 & 0.006 & 0.050
    \end{tabular}
    \caption{Parameters for each trial functional fit to the values of $\mu t_\mathrm{min}$ from $c=0$ data. In addition, we list the RMS errors from comparing the results of simulations with $c=0$, $c=10^{-5}$, and $c=10^{-4}$ with these fits. The logarithmic function provides a much better fit and extrapolation than the square root function.}
    \label{tab:fit_parameters}
\end{table}
In Fig.~\ref{fig:extrapolate} we show the time of the first minimum of $\langle P_z\rangle$ for $c=0$ (black dots), $c=10^{-5}$ (red stars), and $c=10^{-4}$ (gray crosses). In addition we fit $t_\mathrm{min}$ as a function of $N_\mathrm{sites}$ using both a square root function (dotted curve) and a logarithmic function (solid curve). The best fit parameters of each are listed in Tab.~\ref{tab:fit_parameters}. The logarithmic function provides a better fit for the $c=0$ data (indicated by the RMS error $\epsilon_{c=0}$), and a much better extraploation to larger $N_\mathrm{sites}$ (indicated by $\epsilon_{c=10^{-5}}$ and $\epsilon_{c=10^{-4}}$). 

For our chosen conditions, the physical inter-neutrino spacing is approximately $10^{-11}\,\mathrm{cm}$, which would be impossible to simulate directly on the domain sizes required to produce neutrino flavor instabilities. However, it may be possible to resolve the FFI in a MB calculation if the timescale of MB evolution is comparatively slow. A logarithmic fit indicates that $\sim 11,600$ sites are needed for $t_{\mathrm{min}}$ to match the mean-field (MF) FFI growth timescale, which already exceeds the largest MB neutrino simulations to date. However, several e-folding times are required to reach saturation of the MF FFI, demanding a much larger number of sites in MB simulations if the MB effects are to manifest after FFI saturation. If there is one site per neutrino (i.e. $10^{11}$ sites), the projected MB $t_\mathrm{min}=10.9\mu^{-1}$ corresponds to 21.8 $e$-folding times (i.e., 9.4 orders of magnitude). The effective in-medium mixing angle at a density of $10^{12}$ g/cm$^3$ is on the order of $10^{-10}$ ($\approx e^{-23}$) radians. Using this as an optimistic estimate for the magnitude of an initial perturbation, in many astrophysical environments MB effects could occur before the FFI saturates. However, the magnitude of the perturbation would be larger at lower matter densities and stronger inhomogeneities, potentially allowing mean-field effects to take hold first.

\textit{Conclusion --}
In this work, we performed the first systematic treatment of inhomogeneous many-body (MB) dynamics of the Fast Flavor Instability (FFI) to demonstrate that MB effects are expected to persist into thermodynamically large distributions of neutrinos. Our numerical method treats interactions and neutrino advection in a way that systematically approaches instantaneous local interactions in the limit of large system size. In particular, our MB simulations revealed rapid, early flavor transformation, occurring well before MF predictions, combined with kinematic decoherence effects common to inhomogeneous calculations. We observed that as the number of computational sites increases, the flavor transformation dynamics become slower and the entanglement entropy decreases, although the evolution timescales only scale as $\log(N_\mathrm{sites})$. This is consistent with previous work that probed similar homogeneous systems \cite{martin2022classical}. Our work represents the first time the mean-field FFI and previous MB results are all reproducible under a single framework with spatially resolved calculations.

We note that our Hamiltonian is designed to treat only forward scattering, neglecting direction-changing terms, although the expectation that these extra terms increase the rate of onset of MB effects does not contradict the conclusions of this work \cite{cirigliano2024neutrino}. In addition, future calculations should focus in increased number of particles to further test the asymptotic scaling with $N_\mathrm{sites}$ and do so in multiple dimensions. Dimensionality is particularly important, as it is well known that high dimensionality causes many systems to be well described by the mean-field approximation. Our Hamiltonian is similar to the Heisenberg model, for which the mean field is accurate only for $d\geq 4$, although it is unclear if this intuition can be trivially mapped to our system. Furthermore, we note that our study includes $\approx 10^{28}$ neutrinos per site; simulating one neutrino per site is currently computationally infeasible. Finally, although our method imposes inhomogeneity through the shape function term in the Hamiltonian, independent methods of treating inhomogeneity (e.g., wavepackets formalism \cite{akhmedov2017collective,friedland2003many,johns2025local,kost2024once}) should be explored.

\textit{Acknowledgements --} We have benefitted from useful conversations with Baha Balantekin, Joseph Carlson, Michael Cervia, Julien Froustey, Lucas Johns, Gail McLaughlin, Hiroki Nagakura, and Ermal Rrapaj. This material is based upon work supported by the National Science Foundation under Award No. PHY-2412683.

\bibliography{apssamp}

\end{document}